\documentclass[aip,reprint]{revtex4-1}

\usepackage{graphicx}
\usepackage{dcolumn}
\usepackage{bm}

\usepackage{caption}
\usepackage{subcaption}
\usepackage{multirow}
\usepackage{color}

\draft 

\begin{document}


\title{Inner-shell photoionization and core-hole decay of Xe and XeF$_2$} 



\author{Stephen H. Southworth}
\affiliation{X-ray Science Division, Argonne National Laboratory, Argonne, IL 60439, USA}

\author{Ralf Wehlitz}
\affiliation{Synchrotron Radiation Center, University of Wisconsin--Madison, Stoughton, WI 53589, USA}

\author{Antonio Pic\'on}
\affiliation{X-ray Science Division, Argonne National Laboratory, Argonne, IL 60439, USA}

\author{C. Stefan Lehmann}
\affiliation{X-ray Science Division, Argonne National Laboratory, Argonne, IL 60439, USA}

\author{Lan Cheng}
\affiliation{Department of Chemistry and Biochemistry, The University of Texas at Austin, Austin, TX 78712, USA}

\author{John F. Stanton}
\affiliation{Department of Chemistry and Biochemistry, The University of Texas at Austin, Austin, TX 78712, USA}


\date{\today}

\begin{abstract}
Photoionization cross sections and partial ion yields of Xe and XeF$_2$ from Xe 3d$_{5/2}$, Xe 3d$_{3/2}$, and F 1s subshells in  the 660--740 eV range are compared to explore effects of the F ligands.  The Xe 3d - $\epsilon$f continuum shape resonances dominate the photoionization cross sections of both the atom and molecule, but prominent resonances appear in the XeF$_2$ cross section due to nominal excitation of Xe 3d and F 1s electrons to the lowest unoccupied molecular orbital (LUMO), a delocalized anti-bonding MO.  The subshell ionization thresholds, the LUMO resonance energies and their oscillator strengths are calculated by relativistic coupled-cluster methods.  Several charge states and fragment ions are produced from the atom and molecule due to alternative decay pathways from the inner-shell holes.  Total and partial ion yields vary in response to the shape resonances and LUMO resonances.  Previous calculations and measurements of atomic Xe 3d core-hole decay channels and our calculated results for XeF$_2$ guide interpretations of the molecular ion products.    The partial ion yields of XeF$_2$ are dominated by Xe 3d core-hole decays, but distinct ion products are measured at the F 1s - LUMO resonance.  Xe 3d core-hole decays from XeF$_2$ produce lower charge states in comparison with atomic Xe, and energetic F ions are produced by Coulomb explosions of the molecular ions.  The measurements support a model of molecular core-hole decay that begins with a localized hole, stepwise Auger electron emission spreads charge across neighboring atoms, and the system fragments energetically.  
\end{abstract}


\maketitle 


\section{INTRODUCTION}
The xenon fluorides, XeF$_n$ (n = 2, 4, 6), have been the subject of numerous studies that have focused on their structures, physical, chemical, and spectroscopic properties.  The nature of the chemical bonding in ``noble gas compounds'' \cite{grochala2007, liao1998} or ``compounds with an excess of electrons'' \cite{charkin1978} has been of continuing interest.  Despite the closed-shell structure of atomic Xe, the high electron affinities of the F ligands induce bond formation by sharing charge with the Xe valence electrons.  An early bonding model for linear XeF$_2$ suggested 3-center, 4-electron delocalized molecular orbitals (MOs) consisting of 2p$_z$ atomic orbitals (AOs) on the F atoms and the 5p$_z$ AO on Xe, where z is taken as the molecular axis \cite{rundle1963, wilson1963}.  In this model, two electrons are in a bonding orbital delocalized over the three centers, and two electrons are in a nonbonding orbital on the F atoms only.  The resulting picture is of a weakly-bonded system with mixed ionic and covalent character.  The third, unoccupied 3-center orbital is the anti-bonding combination and was identified as the lowest unoccupied molecular orbital (LUMO) in photoabsorption spectra. \cite{wilson1963, comes1973, nielsen1976, basch1971}

The Hartree-Fock self-consistent-field (SCF) calculations of Basch \textit{et al.} \cite{basch1971} provide a more comprehensive model of bonding in the xenon fluorides and of their physical properties.  Population analysis of the AO composition of the MOs, the atomic charge distributions, and the orbital energies derived from the SCF calculations are compared with electron spin resonance, nuclear magnetic resonance, photoabsorption, and photoelectron measurements. \cite{basch1971}  For example, core-level binding energy shifts measured in x-ray photoelectron spectra are sensitive to the valence electron charges on the Xe and F atoms.\cite{basch1971,carroll1974}  More recent quantum chemical calculations treat relativistic and many-electron interactions in the xenon fluorides. \cite{styszynski1997, buth2003a, pernpointner2005}

\begin{figure}
\begin{center}
\includegraphics [trim=3cm 7cm 3cm 7cm,clip,width=8cm]{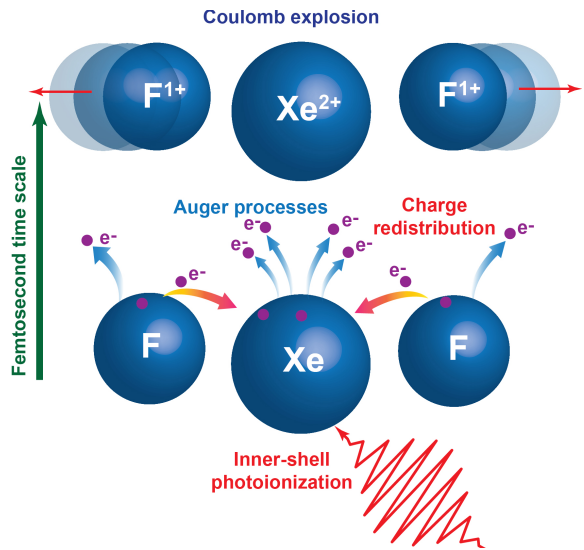}
\caption{\small Illustration of inner-shell photoionization, core-hole decay, charge redistribution, and Coulomb explosion in XeF$_2$.  Inner-shell photoionization of Xe triggers core-hole decay that is initially localized but subsequently involves delocalized valence electrons.  Charge spreads to neighboring atoms and the system Coulomb explodes.  The processes proceed concurrently on the femtosecond time scale.}
\label{XeF2_sketch}
\end{center}
\end{figure}

Here we explore molecular effects on inner-shell photoionization and core-hole decay by comparing measurements and calculations on XeF$_2$ with atomic Xe.  Buth \textit{et al.} \cite{buth2003b} explained how the observed increase in the decay width of Xe 4d holes with the number of F ligands results from interatomic electronic decay processes.  Dunford \textit{et al.} \cite{dunford2012} explored charge production, charge redistribution, and energetic ion fragmentation in the decay of Xe 1s holes in XeF$_2$.  The initial core hole is highly localized, but as the core hole decays by the emission of fluorescent photons and Auger electrons, holes are formed in the outer subshells, charge is redistributed to the F ligands, and the system undergoes Coulomb explosion.  The kinetic energies of the F fragment ions measured by Dunford \textit{et al.} \cite{dunford2012} are smaller than the Coulomb energies that would be stored in the system at its ground state geometry.  This suggests that ion fragmentation begins while the charge distribution is developing on the system.  The picture that emerges is sketched in Fig. \ref{XeF2_sketch} with core-hole decay, charge redistribution, and Coulomb explosion proceeding concurrently on the femtosecond time scale.

Inner-shell photoionization of molecules is also being explored with intense, ultrafast x-ray free-electron laser (XFEL) pulses that induce absorption of multiple x-ray photons.\cite{fang2012,erk2013}  In a recent pump-probe experiment on iodomethane, an optical laser dissociated the I atom and XFEL pulses triggered inner-shell photoionization and core-hole decay as a function of time delay.\cite{erk2014}  Charge transfer between the I and CH$_3$ fragments was measured as a function of internuclear separation, and the results are consistent with a classical over-the-barrier model of electron transfer.  Such experiments provide new insights into molecular core-hole decay.

In the present work, we used soft x rays in the 660--740 eV range to photoionize the Xe 3d$_{5/2}$, Xe 3d$_{3/2}$, and F 1s subshells of Xe and XeF$_2$.  The Xe 3d and F 1s ionization energies are similar, and the photoionization cross sections overlap, but the inner-shell states are atomic-like, so the initial core holes are localized.  As Auger electron emission proceeds stepwise, delocalized valence electrons participate, charge is spread across the molecule, and the system fragments.  An ion spectrometer was used to measure total and partial ion yields across the subshell edges.  The 3d photoionization cross section of atomic Xe was measured previously, \cite{becker1987,saito1992,arp1999,kivimaki2000,kato2007} and variations of the yields of the multicharged ions Xe$^{q+}$ (q = 1--8) vs. photon energy were reported.\cite{saito1992}  The atomic Xe 3d core-hole decay processes have been characterized using Auger electron spectroscopy, calculated de-excitation pathways, and electron-ion and electron-electron coincidence spectroscopies.\cite{tamenori2002,jonauskas2003,partanen2005,viefhaus2005,suzuki2011}  Our goal is to explore the effects of the F ligands on the photoionization cross sections and partial ion yields of XeF$_2$.  The Xe 3d - $\epsilon$f continuum shape resonances dominate the 3d$_{5/2}$ and 3d$_{3/2}$ photoionization cross sections of both the atom and molecule, but distinct resonances are also observed in XeF$_2$ due to excitation of the LUMO from the Xe 3d$_{5/2}$, Xe 3d$_{3/2}$, and F 1s subshells.  The LUMO is a delocalized anti-bonding MO that was identified in early far-UV photoabsorption measurements \cite{wilson1963,comes1973,nielsen1976,basch1971} and contributed to understanding the electronic structure of XeF$_2$ at the qualitative molecular orbital level.  Theoretical treatments of inner-shell photoionization and core-hole decays in molecules are at an early stage compared with atoms.  However, we performed relativistic coupled-cluster calculations to determine the subshell ionization energies and the energies and oscillator strengths of the LUMO resonances.  The calculations help guide interpretations of the ion yields.

Section \ref{methods} of this paper describes the ion yield and branching ratio measurements, and computational methods are discussed in Section \ref{theory}.  In Section \ref{results} we discuss the measured and calculated results for ion yields, branching ratios, energies and oscillator strengths.  Conclusions and suggestions for future research are in Sec. \ref{conclusion}.


\section{EXPERIMENTAL METHODS}
\label{methods}

\begin{figure}
\begin{center}
\includegraphics [trim=6cm 6cm 7cm 5cm,clip,width=8cm]{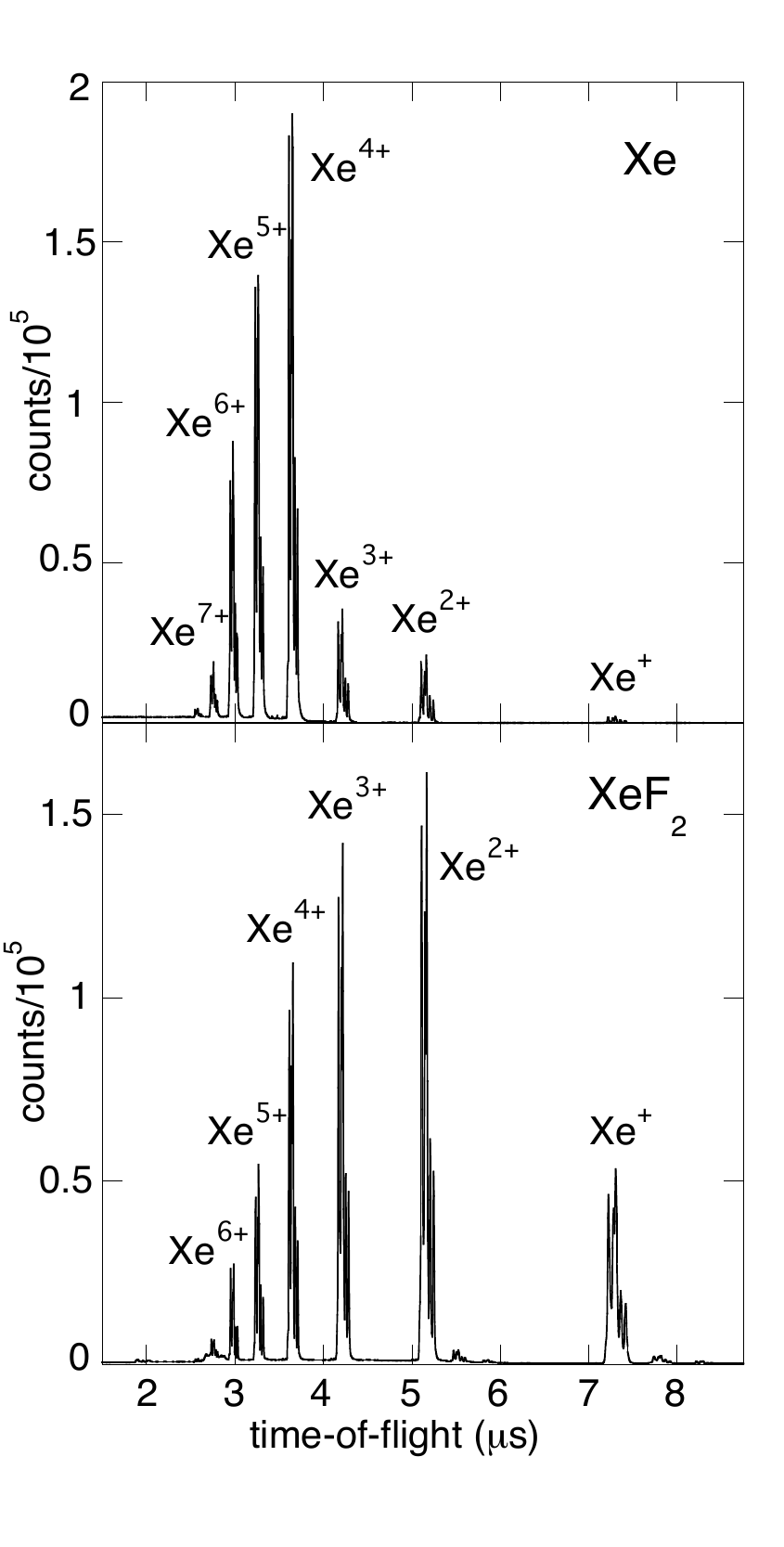}
\caption{\small Ion time-of-flight spectra of Xe (top) and XeF$_2$ (bottom) summed over the 695--730 eV photon energy range.  The structures in the peaks are due to the different Xe isotope masses.  The weaker peaks of the XeF$_2$ spectrum are shown on an expanded scale in Fig. \ref{XeF2_iTOF_X}.}
\label{Xe_XeF2_iTOF}
\end{center}
\end{figure}

Total and partial ion yields of Xe and XeF$_2$ were measured following photoionization of the Xe 3d$_{5/2}$, Xe 3d$_{3/2}$, and F 1s subshells in the 660--740 eV range.  An ion time-of-flight spectrometer with pulsed extraction field\cite{wehlitz2002,hartman2013} was used on the variable line spacing plane grating monochromator (VLS-PGM) beamline at Wisconsin's Synchrotron Radiation Center.\cite{reininger2005}  Xe was taken from a compressed gas cylinder, and the 5 mbar vapor pressure of XeF$_2$ was used from the solid at room temperature.  Diffuse beams of the samples were produced with a capillary nozzle in the vacuum chamber and crossed with the photon beam in the interaction volume of the ion spectrometer.  The ion spectrometer uses a pulsed electric field to push ions from the interaction region through an aperture into the time-of-flight drift tube for mass/charge dispersion and single-ion detection with a microchannel plate detector.  The ion time-of-flight spectra from Xe and XeF$_2$ recorded over the 695--730 eV range are summed and plotted in Fig. \ref{Xe_XeF2_iTOF}.  The ion spectra in this photon energy range are dominated by Xe 3d core hole decays.  The structures in the ion peaks are from the different masses and natural abundances of the Xe isotopes.  The isotopically averaged mass of Xe is 131.3 while F has a single isotope with mass 19.0.

\begin{figure}
\begin{center}
\includegraphics [trim= 6cm 9cm 7cm 10cm,clip,width=8cm]{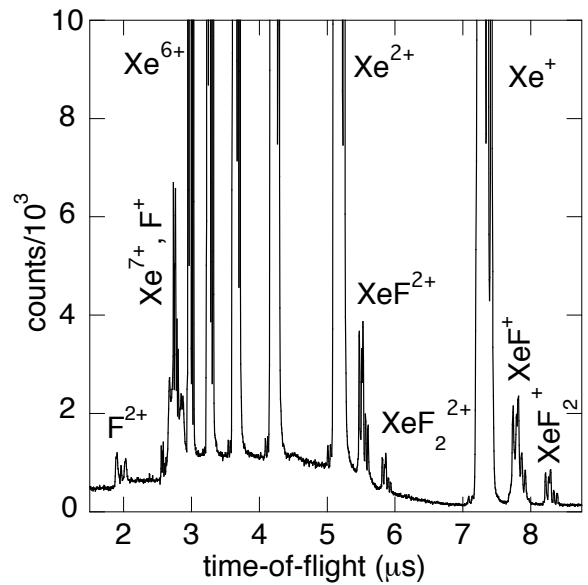}
\caption{\small Ion time-of-flight spectrum of XeF$_2$ on an expanded scale to show the weaker peaks.  The F$^+$ and Xe$^{7+}$ peaks overlap.  The ion spectrometer has small detection efficiencies for energetic F$^+$ and F$^{2+}$ fragment ions.  The F$^+$ and F$^{2+}$ spectra are also plotted in Fig. \ref{F_F2_iTOF}.}
\label{XeF2_iTOF_X}
\end{center}
\end{figure}

\begin{figure}
\begin{center}
\includegraphics [trim= 6cm 6cm 7cm 5cm,clip,width=8cm]{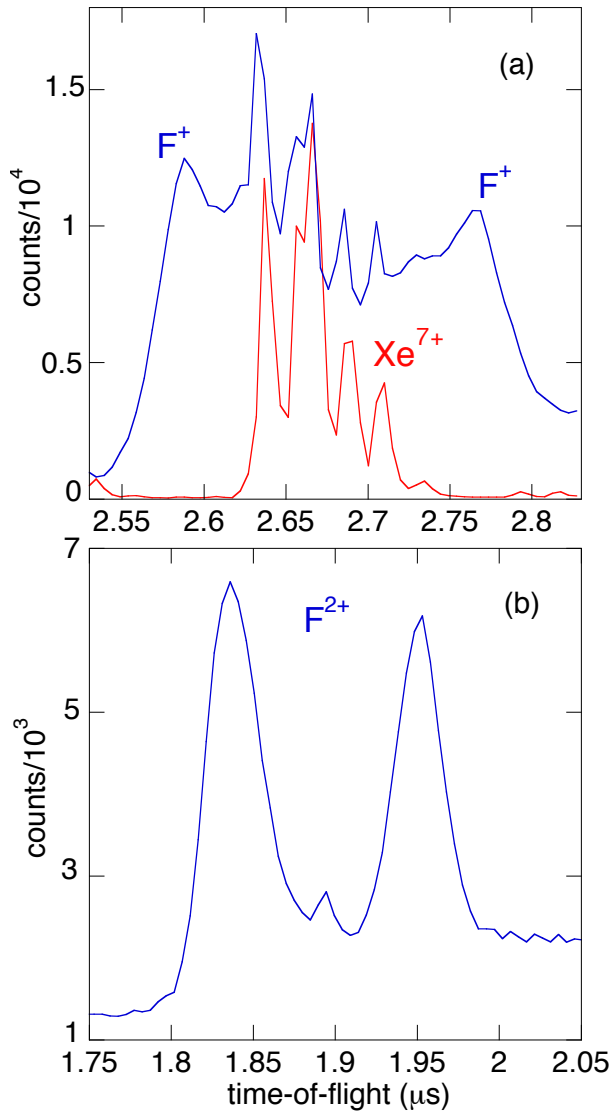}
\caption{\small Ion spectra of F$^+$ and F$^{2+}$ from photoionization of XeF$_2$ summed over the 695--730 eV photon energy range.  (a) F$^+$ overlapped with Xe$^{7+}$ (blue curve) compared with Xe$^{7+}$ from atomic Xe (red curve).  The F$^+$ component is split due to $\sim$15 eV kinetic energy.  (b) F$^{2+}$ peak splitting corresponds to $\sim$27 eV kinetic energy.}
\label{F_F2_iTOF}
\end{center}
\end{figure}

The ion spectrum of XeF$_2$ is plotted on an expanded scale in Fig. \ref{XeF2_iTOF_X} to show the weaker peaks.  Strong Xe$^{q+}$ (q = 1--6) peaks and weak XeF$_2^+$, XeF$_2^{2+}$, XeF$^+$, and XeF$^{2+}$ peaks were recorded, but the F$^+$ (overlapped with Xe$^{7+}$) and F$^{2+}$ peaks are small.  This is attributed to the low detection efficiency of the spectrometer for energetic ions produced by Coulomb explosions.  The spectrometer is designed to measure the charge-state yields of atoms and parent molecular ions that have only small thermal energies.\cite{wehlitz2002,hartman2013}  Energetic fragment ions leave the interaction volume during delays between extraction pulses so most are not detected.  The collection solid angles for energetic ions are limited to a small range along the symmetry axis of the spectrometer.  The F$^+$ and F$^{2+}$ peaks measured over 695--730 eV are plotted in Fig. \ref{F_F2_iTOF}.  The F$^+$ and Xe$^{7+}$ peaks overlap as shown in Fig. \ref{F_F2_iTOF}(a) by comparing with the Xe$^{7+}$ spectrum from atomic Xe.  The F$^+$ ions are spread to higher and lower flight times than the Xe$^{7+}$ ions due to their momenta from fragmentation of molecular ions.  The average kinetic energy of F$^+$ is estimated to be $\sim$15 eV from the peak splitting and extraction field of the spectrometer.  The splitting of the F$^{2+}$ peaks in Fig. \ref{F_F2_iTOF}(b) gives an average kinetic energy of $\sim$27 eV.  Although energetic ions are not efficiently detected with our spectrometer, variations of the measured yields of F$^+$ and F$^{2+}$ give useful information and are discussed in Sec. \ref{results}.

Figure \ref{Xe_XeF2_iTOF} shows that the Xe charge state distribution shifts down in the molecule, which we attribute to redistribution of charge to the F ligands.  For atomic Xe, the average charge over the 695--730 eV range is 4.5, while the average Xe charge for XeF$_2$ is 2.6 (see Table \ref{Tab:charge-ratios}).  We infer that, on average, two charges are redistributed to the F ligands.  The parent ions, XeF$_2^+$ and XeF$_2^{2+}$, and the ions missing one F ligand, XeF$^+$ and XeF$^{2+}$, are weak compared with Xe$^{q+}$ (q = 1--6).  This suggests that most of the molecular ions dissociate to three atomic ions.  Since XeF$_2$ is a linear molecule and the two F ligands have the same mass, both much lighter than Xe, momentum conservation shows that the Xe ions are produced with near zero velocities.  Also, the Xe ion spectra show distinct isotope structure without broadening that would result from kinetic energy released in the fragmentation process.  The Xe fragment ions are therefore efficiently extracted and detected, and the Xe$^{q+}$ charge state distributions from the atom and molecule can be compared.

For inner-shell photoionization, total ion yields vs. photon energy are proportional to photoabsorption cross sections to a good approximation and the fluorescence yield is small in our case.  To measure total ion yields in this experiment, the pulsed extraction field was replaced with a static field that pushed all of the ions, including energetic fragment ions, into the spectrometer.  The monochromator was calibrated by recording ion yields of Ne across the 1s - 3p resonance at 867.12 eV \cite{coreno_Ne} and O$_2$ across the 1s - $^{3}\Pi_u$ resonance at 530.8 eV.\cite{coreno_O2}  The monochromator resolution was $\sim$0.7 eV at 700 eV.  Ion spectra were recorded over $\sim$660--740 eV with steps of 0.5 or 1 eV.  For each photon energy, the counts under each ion peak were summed and divided by the total counts to determine branching ratios.  The total and partial ion yields vs. photon energy are discussed in Sec. \ref{results}.


\section{THEORY}
\label{theory}
In this section we present the method that we used to calculate the inner-shell excitations and binding energies in XeF$_2$ at the Xe 3d and F 1s edges. In our particular case, the goal was to calculate the transition energy of Xe 3d$_{5/2}$ - LUMO, Xe 3d$_{3/2}$ - LUMO, and F 1s - LUMO, together with the corresponding binding energy for Xe 3d$_{5/2}$, Xe 3d$_{5/2}$, and F 1s. Molecular inner-shell calculations are demanding; they require a high-degree treatment of both electron-correlation and orbital-relaxation effects, as strong relaxation of the wave function is induced by the removal of an inner-shell electron. In addition, relativistic effects are more significant for inner-shell electrons than for valence electrons, since the former are in general moving faster. Here the explicit calculation of inner-shell electrons necessitates the use of all-electron relativistic methods. Therefore, in our calculations we have employed all-electron relativistic coupled cluster (CC) approaches in order to obtain a systematic treatment aiming at high accuracy.

CC \cite{CCgradbook} is a powerful numerical method used in quantum chemistry that utilizes an exponential parametrization of the wave function in order to provide a rapidly convergent and size-extensive description for energy and properties. CC provides very good results for single-reference systems, whose electronic wave functions are dominated by a leading electronic configuration. In the CC framework, the calculations for excited or ionized states are based on the equation-of-motion (EOM) CC methods (also known as CC linear response theory in the literature).\cite{cclrt1,cclrt2,cclrt3,cclrt4,cclrt5,cclrt6} 
EOMCC is a ``direct'' approach that computes the difference between a target state (excited or ionized state) and a reference state (usually chosen as the closed-shell ground state), and thus can effectively exploit the cancelation of errors in the description of the two states. EOMCC calculations for core excitations or ionizations have been reported in the literature. \cite{Bartlett1995,Besley2012, Coriani2012} As it is the spin-orbit coupling that introduces high computational overheads for the fully relativistic CC calculations, in our present calculations we have adopted the cost-effective scheme of combining scalar-relativistic EOMCC calculations with spin-orbit corrections obtained at the Hartree-Fock (HF) level. Specifically, we have employed the spin-free exact two-component theory in its one-electron variant (SFX2C-1e) \cite{DyallNESC4, XQR2009} in combination with EOMCC methods, in order to achieve an efficient and accurate treatment of scalar-relativistic effects as well as a systematic incorporation of electron-correlation and orbital-relaxation effects. The SFX2C-1e EOMCC results for the excitation and binding energies are then augmented with the spin-orbit contributions obtained at the level of Koopmans' theorem using the Dirac-Coulomb-Gaunt (DCG) Hamiltonian. \cite{Dyallbook} The non-relativistic HF, SFX2C-1e HF, and SFX2C-1e EOMCC calculations have been performed using CFOUR, \cite{CFOUR, eomip, EOMCCSDgrad1, EOMCCSDgrad2, EOMCCSDT3, Cheng2011}while the DC and DCG HF calculations have been carried out using the DIRAC program package. \cite{DIRAC, SaueDHF1, SaueDHF2} 

In Table \ref{Tab:Lan_Results1} we show the results for the binding energies calculated for Xe atom, HF molecule, and XeF$_2$ molecule. The equilibrium geometry calculated at the SFX2C-1e/CCSD(T)/unc-ANO-RCC level (r$_{\text{Xe-F}}=1.9736$ \AA) \cite{Cheng2013} has been used for the XeF$_2$ molecule, while the experimental bond length (r$_{\text{H-F}}$=0.9168 \AA) \cite{Herzberg} has been used for the HF molecule. First, we present results obtained by Koopmans' theorem using the non-relativistic, SFX2C-1e, DC, and DCG Hamiltonians. The ANO-RCC basis in its uncontracted form (unc-ANO-RCC) \cite{faegribas, Roos2004} has been used in these calculations. The importance of relativistic effects is demonstrated by the difference between the non-relativistic and SFX2C-1e results, while the spin-orbit corrections are obtained as the difference between the SFX2C-1e results and those from DC and DCG calculations. In order to account for electron-correlation and orbital-relaxation effects, we perform several calculations at the SFX2C-1e-EOM-CCSD and SFX2C-1e-EOM-CCSDT \cite{EOMCCSDT2} levels. Systematically enlarged atomic-natural-orbital basis sets of double-zeta, triple-zeta, and quadruple-zeta quality (ANO0, ANO1, and ANO2) \cite{anox2c} as well as the unc-ANO-RCC basis have been used to demonstrate the basis-set convergence. In all the CC calculations, the Xe 1s, 2s, and 2p orbitals have been frozen. Here it is interesting to observe the atomic nature for the inner-shell molecular orbitals: the correction energies for the Xe 3d (or the F 1s) orbitals are similar in the Xe (or the F) atom than in the Xe (or the F) site of the XeF$_2$ molecule. The total binding energy has been obtained by combining the results of the best quality of accounting for various effects, including the DCG/unc-ANO-RCC results at the level of Koopmans' theorem, the electron-correlation and orbital-relaxation corrections at the EOM-CCSD/unc-ANO-RCC level, and the triples corrections taken as the difference between the EOM-CCSDT/ANO1 and EOM-CCSD/ANO1 results.

The positions of the LUMO resonance have been obtained by an additive scheme, by augmenting the accurately determined binding energies with the difference between the excitation and binding energies calculated at the SFX2C-1e-EOM-CCSD/ANO1 level (see the last row of Table \ref{Tab:Lan_Results1}). In Table \ref{Tab:Lan_Results2} we also show the calculated oscillator strengths for the corresponding resonant transitions in XeF$_2$. We mention that, due to the D$_{\text{2h}}$ symmetry of the XeF$_2$ molecule as well as the atomic nature of the inner-shell orbitals, the atomic Xe 3d orbitals almost exactly map onto the corresponding molecular orbitals. Therefore we have used the atomic notation for the Xe 3d orbitals in \ref{Tab:Lan_Results2}. The calculated and measured results are compared in Sec. \ref{results}.

Convergence problems in the solution of the EOMCC eigenvalue equations may occur for the standard Davidson algorithm, when there exist multiply excited or ``shake-up'' configurations that are energetically quasi-degenerate with the target inner-shell excited or ionized states. Among our calculations, convergence difficulties have been encountered in the EOM-CCSD/unc-ANO-RCC and all the EOM-CCSDT calculations of the binding energies as well as the EOM-CCSD/ANO1 calculations of excitation energies. Here we have adopted the Arnoldi algorithm \cite{arnoldi} to converge the solutions for these calculations. In the EOM-CCSD/unc-ANO-RCC calculations, virtual orbitals with orbital energies higher than 100 Hartree have also been excluded to expedite the convergence. \cite{footnote2}

\begin{widetext}
\begin{table}[h]
\resizebox{0.8\textwidth}{!}{\begin{minipage}{\textwidth}
\begin{tabular}{ |l|c|c|c|c|c|c|c| }
\hline
&\multicolumn{2}{ c| }{Xe 3d in Xe atom} & F 1s in HF & &  \multicolumn{3}{ c| }{XeF$_2$ molecule} \\ \hline
& 3d$_{3/2}$ & 3d$_{5/2}$ & F 1s & & Xe 3d$_{3/2}$ & Xe 3d$_{5/2}$ & F 1s \\ \hline
\multicolumn{8}{ |c| } {Koopmans' theorem} \\ \hline
Nonrel/unc-ANO-RCC   & 710.7 & 710.7 & 715.5 & & 714.4 & 714.4 & 715.0 \\
SFX2C-1e/unc-ANO-RCC & 700.2 & 700.2 & 716.2 & & 703.9 & 703.9 & 715.6 \\
DC/unc-ANO-RCC       & 708.1 & 694.9 & 716.3 & & 711.9 & 698.6 & 715.7 \\
DCG/unc-ANO-RCC      & 707.5 & 694.6 & 716.1 & & 711.2 & 698.3 & 715.4 \\ \hline
\multicolumn{8}{ |c| } {SFX2C-1e-EOM-CCSD corrections} \\ \hline
ANO0                 & -10.9 & -10.9 & -19.2 & & -11.5 & -11.5 & -20.2 \\
ANO1                 & -16.3 & -16.3 & -20.0 & & -16.7 & -16.7 & -21.0 \\
ANO2                 & -17.6 & -17.6 & -20.3 & & -17.9 & -17.9 & -21.3 \\ 
unc-ANO-RCC          & -17.9 & -17.9 & -19.9 & & -18.0 & -18.0 & -21.2 \\ \hline
\multicolumn{8}{ |c| } {EOM-CCSDT corrections on top of EOM-CCSD} \\ \hline
ANO0                 & -0.9  & -0.9  & -1.7  & & -1.2  & -1.2  & -2.9 \\
ANO1                 & -0.7  & -0.7  & -1.8  & & -1.0  & -1.0  & -2.9 \\ \hline
Total Binding Energy & 688.9 & 676.0 & 694.4 & & 692.2 & 679.3 & 691.3 \\ 
LUMO resonance       &       &       &       & & 682.6 & 669.7 & 683.1 \\  
\hline
\end{tabular}
\caption{Computed binding energies and LUMO resonance positions (in eV).}
\label{Tab:Lan_Results1}
\end{minipage} }
\end{table}
\end{widetext}

\begin{widetext}
\begin{table}[h]
\resizebox{0.8\textwidth}{!}{\begin{minipage}{\textwidth}
\begin{tabular}{ |l|c|c|c|c|c|c|c| }
\hline
&\multicolumn{2}{ c| }{F 1s - LUMO} &  \multicolumn{5}{ c| }{Xe 3d - LUMO} \\ \hline
& 1s(F\#1)+1s(F\#2) & 1s(F\#1)-1s(F\#2) & 3d(2z$^2$-x$^2$-y$^2$) & 3d(xz) & 3d(yz) & 3d(xy) & 3d(x$^2$-y$^2$) \\ \hline
ANO0 & 0.0627 & 0 & 0.0048 & 0.0013 & 0.0013 & 0 & 0 \\
ANO1 & 0.0611 & 0 & 0.0047 & 0.0013 & 0.0013 & 0  & 0 \\ 
\hline
\end{tabular}
\caption{Oscillator strength for the F 1s to LUMO and the Xe 3d to LUMO transitions in the XeF$_2$ molecule computed at the EOM-CCSD level.}
\label{Tab:Lan_Results2}
\end{minipage} }
\end{table}
\end{widetext}


%
%

%

\section{RESULTS}
\label{results}

\subsection{Total cross sections}
\begin{figure}
\begin{center}
\includegraphics [trim=6cm 9cm 7cm 3cm,clip,width=8cm]{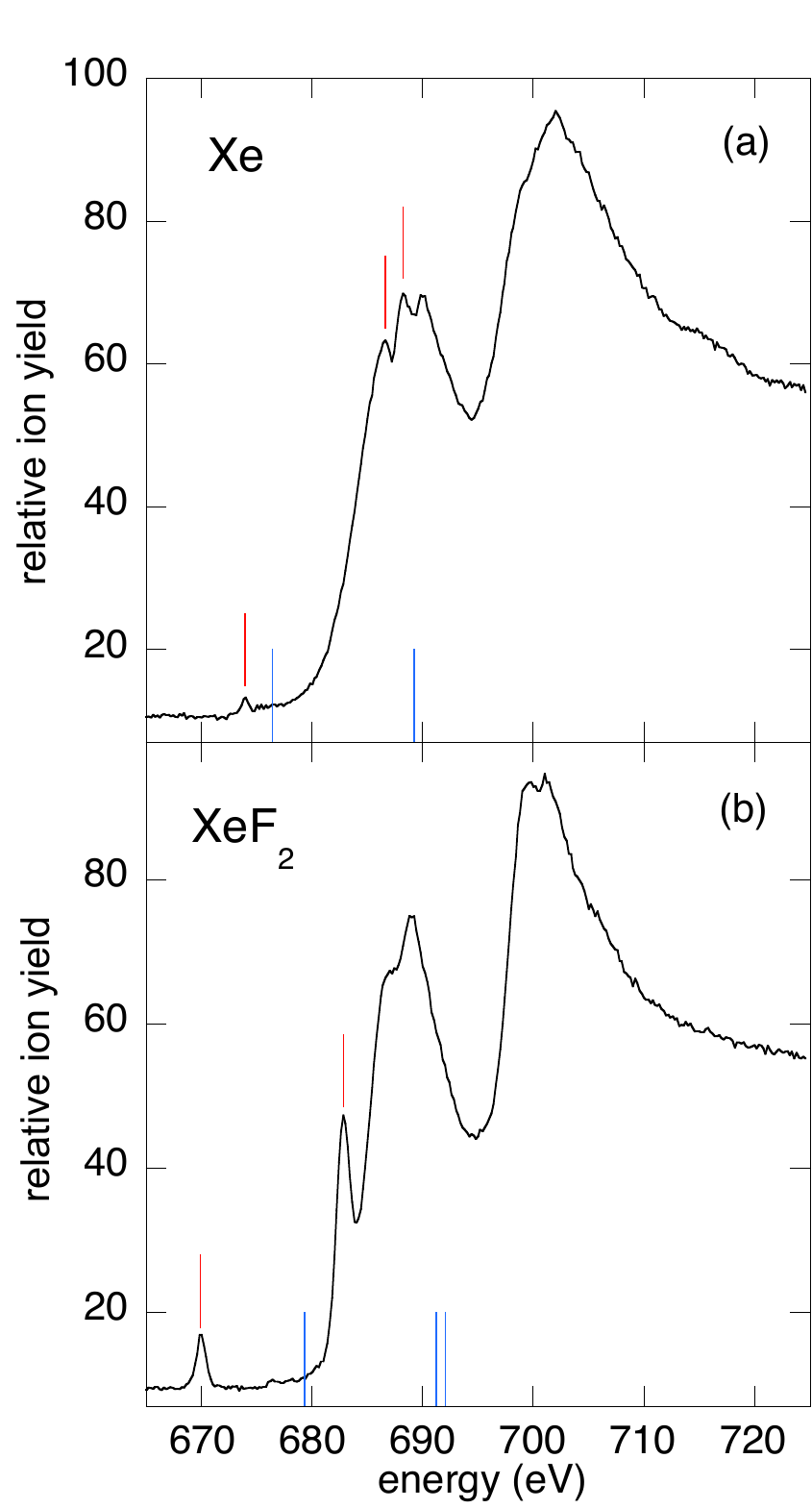}
\caption{\small Total ion yield scans of Xe and XeF$_2$.  Panel (a) is the Xe scan with Rydberg resonances marked by red lines: 3d$_{5/2}$ - 6p (673.9 eV), 3d$_{3/2}$ - 6p (686.6 eV), and 3d$_{3/2}$ - 7p (688.2 eV).  The blue lines mark the 3d$_{5/2}$ (676.44 eV) and 3d$_{3/2}$ (689.23 eV) binding energies.\cite{carroll1974}  The strong, broad peaks are the Xe 3d$_{5/2}$ - $\epsilon$f ($\sim$688 eV) and Xe 3d$_{3/2}$ - $\epsilon$f ($\sim$702 eV) shape resonances.  Panel (b) is the XeF$_2$ scan with resonances marked by red lines: Xe 3d$_{5/2}$ - LUMO (669.9 eV) and the unresolved Xe 3d$_{3/2}$ - LUMO and F 1s - LUMO resonances (682.8 eV).  The blue lines mark the Xe 3d$_{5/2}$ (679.31 eV), Xe 3d$_{3/2}$ (692.09 eV), and F 1s (691.23 eV) binding energies.\cite{carroll1974}  The strong, broad peaks are the Xe 3d$_{5/2}$ - $\epsilon$f ($\sim$688 eV) and Xe 3d$_{3/2}$ - $\epsilon$f ($\sim$700 eV) shape resonances.}
\label{Xe_XeF2_XS}
\end{center}
\end{figure}

The total ion yield scans of Xe and XeF$_2$ are plotted in Fig. \ref{Xe_XeF2_XS}.  Our measurements of the Xe 3d cross sections of atomic Xe agree well with previous measurements,\cite{saito1992,arp1999,kivimaki2000,kato2007} but the XeF$_2$ cross section has not been reported previously.  Carroll \textit{et al.} \cite{carroll1974} measured the binding energies of the Xe 3d$_{5/2}$, Xe 3d$_{3/2}$ and F 1s subshells of Xe and XeF$_2$, and our calculated binding energies match their measurements quite well (see Table \ref{table_energies}). The nearly perfect agreement between the computed and experimental numbers can partially be attributed to cancellation of remaining errors in the calculation, as the residual basis set error for the EOM-CCSD correlation correction and that for the triples correction appear to have different sign following the trend seen in Table \ref{Tab:Lan_Results1}. The Xe 3d$_{5/2}$ - 3d$_{3/2}$ spin-orbit splitting is 12.8 eV in both Xe and XeF$_{2}$, but the binding energies are shifted up by 2.9 eV in the molecule.  The F 1s binding energy of XeF$_2$ is shifted down by 5.5 eV relative to that of F$_2$, which places it only 0.9 eV below the Xe 3d$_{3/2}$ binding energy.  The core-level binding energy shifts in XeF$_2$ are sensitive to the valence electron charge distributions on the Xe and F sites.\cite{basch1971,carroll1974}

The atomic Xe 3d$_{5/2}$ and 3d$_{3/2}$ cross sections are dominated by strong, broad maxima near 688 eV and 702 eV that are attributed to shape resonances in the f-wave continuum channels.  The shape resonances can be understood as independent-electron effects, (that is, those that can be understood in the orbital model of electronic structure) but measurements\cite{kivimaki2000} and calculations\cite{amusia2002,toffoli2002,radojevic2003} show coupling between the 3d$_{5/2}$ - $\epsilon$f and 3d$_{3/2}$ - $\epsilon$f channels.  Small features due to 6p and 7p Rydberg states were also recorded in the atomic Xe cross section in agreement with previous measurements.\cite{arp1999,kato2007}

The Xe 3d - $\epsilon$f shape resonances produce strong maxima near 688 eV and 700 eV in  the XeF$_2$ cross sections, but distinct resonances also appear due to excitation of the LUMO.  The Xe 3d$_{5/2}$ - LUMO resonance is at 669.9 $\pm$ 0.3 eV, which is 9.4 eV below the 3d$_{5/2}$ binding energy and is in nearly exact agreement with theory.  The calculated energies of the Xe 3d$_{3/2}$ - LUMO and F 1s - LUMO resonances are within 0.3 eV of each other, which is less than our 0.7 eV resolution, and we recorded a single peak at 682.8 $\pm$ 0.3 eV.  However, the calculated oscillator strength of the F 1s - LUMO resonance is $\sim20\times$ larger than that of the Xe 3d$_{3/2}$ - LUMO resonance (see Table \ref{Tab:Lan_Results2}).\cite{footnote1} Excitation of the LUMO was observed previously in far-UV photoabsorption spectra,\cite{wilson1963,nielsen1976} in photoexcitation of the Xe 4d subshells, \cite{comes1973} and in the Xe \textit{K}-shell.\cite{dunford2012}

\begin{table}
\caption[]{\label{table_energies} Binding energies and excited state energies from theory and experiment.}
\label{Tab:energetics}
\begin{ruledtabular}
\begin{tabular}{cccc}\
target & transition & theory$^c$ (eV) & experiment (eV)\\ \hline
Xe & 3d$_{5/2}$$^{-1}$ & 676.0 & 676.44 $\pm$ 0.05$^a$\\
Xe & 3d$_{3/2}$$^{-1}$ & 688.9 & 689.23 $\pm$ 0.05$^a$\\
HF & F 1s$^{-1}$ & 694.4 & 694.25 $\pm$ 0.08$^b$\\
XeF$_2$ & 3d$_{5/2}$$^{-1}$ & 679.3 & 679.31 $\pm$ 0.05$^a$\\
XeF$_2$ & 3d$_{3/2}$$^{-1}$ & 692.2 & 692.09 $\pm$ 0.05$^a$\\
XeF$_2$ & F 1s$^{-1}$ & 691.3 & 691.23 $\pm$ 0.05$^a$\\
XeF$_2$ & 3d$_{5/2}$ - LUMO & 669.7 & 669.9 $\pm$ 0.3$^c$\\
XeF$_2$ & 3d$_{3/2}$ - LUMO & 682.6 & 682.8 $\pm$ 0.3$^c$\\
XeF$_2$ & F 1s - LUMO & 683.1 & 682.8 $\pm$ 0.3$^c$\\
\end{tabular}
\end{ruledtabular}
$^a$Ref. \citenum{carroll1974}, $^b$Ref. \citenum{shaw1975}, $^c$This work.
\end{table}

\subsection{Xe ion yields}
\begin{figure}
\begin{center}
\includegraphics [trim=5cm 7cm 7cm 3cm,clip,width=8cm]{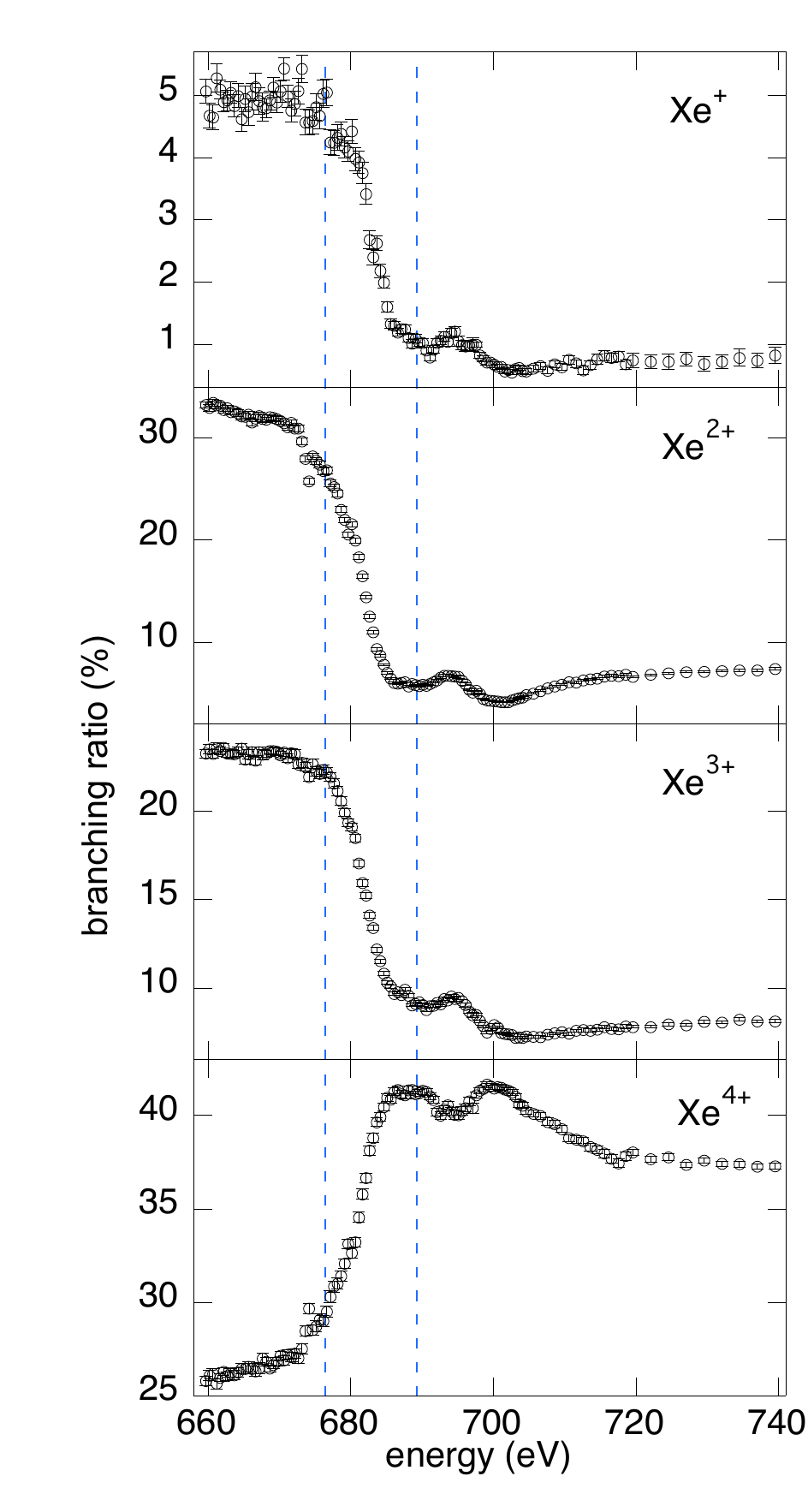}
\caption{\small Branching ratios for Xe$^{q+}$ (q = 1--4) from photoionization of Xe.  The vertical lines mark the 3d$_{5/2}$ (676.44 eV) and 3d$_{3/2}$ (689.23 eV) binding energies.}
\label{Xe_BR_1}
\end{center}
\end{figure}

\begin{figure}
\begin{center}
\includegraphics [trim=5cm 7cm 7cm 3cm,clip,width=8cm]{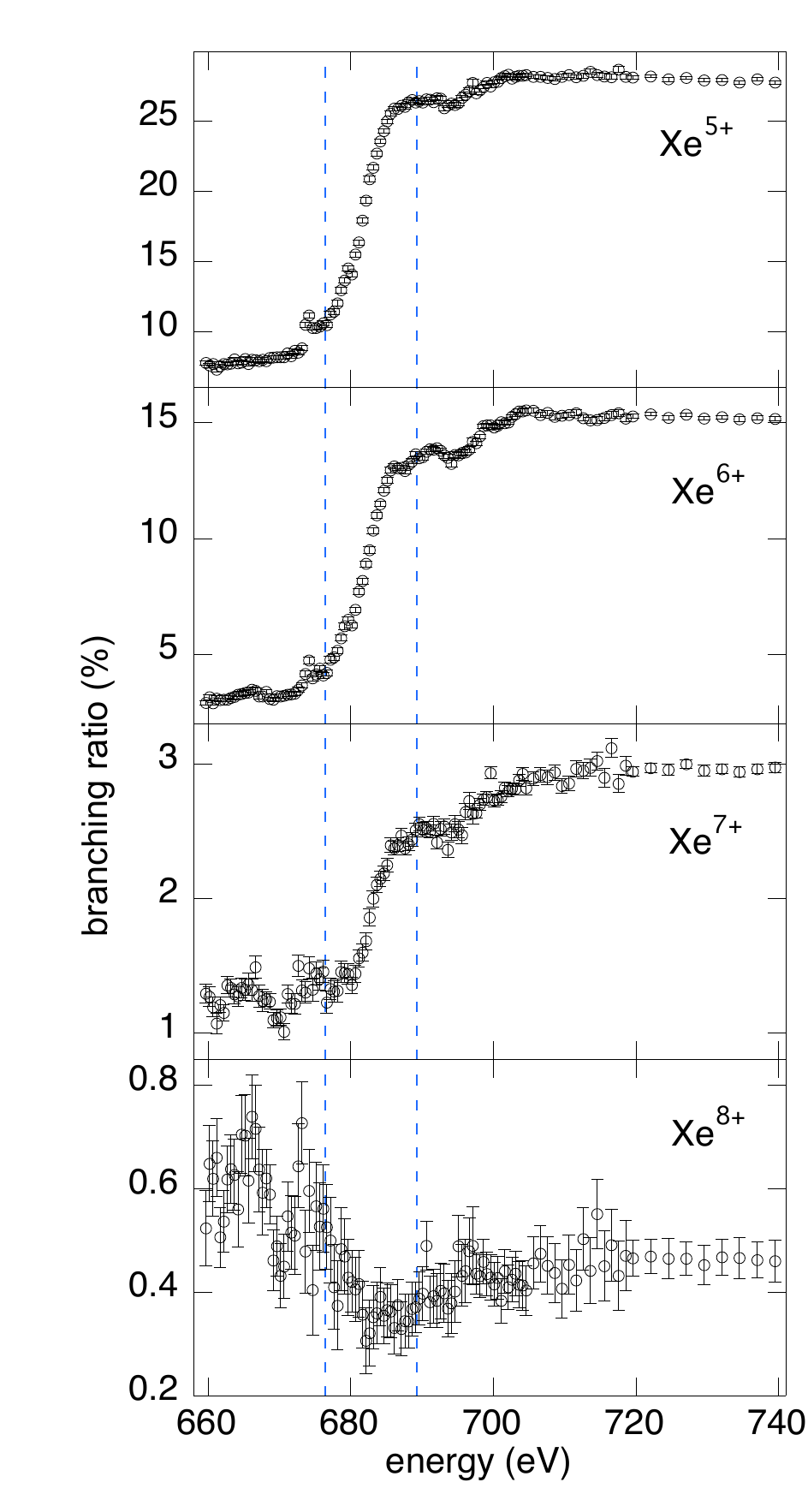}
\caption{\small Branching ratios for Xe$^{q+}$ (q = 5--8) from photoionization of Xe.  The vertical lines mark the 3d$_{5/2}$ (676.44 eV) and 3d$_{3/2}$ (689.23 eV) binding energies.}
\label{Xe_BR_2}
\end{center}
\end{figure}

Ion spectra of atomic Xe were measured over 660--740 eV and the Xe$^{q+}$ (q = 1--8) branching ratios are plotted in Figs. \ref{Xe_BR_1} and \ref{Xe_BR_2}.  The branching ratios strongly change as the photon energy increases from below to above the 3d$_{5/2}$ ionization energy, particularly in response to the $\epsilon$f shape resonance.  The relative yields of the lower charge states, q = 1--3, decrease while q = 4--7 increase.  Higher charge states are expected, as photon energies access deeper subshells and open additional core-hole decay steps.\cite{kochur1994}  Smaller changes in the branching ratios appear as the photon energy increases above the 3d$_{3/2}$ ionization energy and excites its shape resonance.  A second maximum also appears in the 3d$_{5/2}$ cross section at $\sim$707 eV due to coupling with the 3d$_{3/2}$ shape resonance.\cite{kivimaki2000,amusia2002,toffoli2002,radojevic2003}

The present measurements of Xe$^{q+}$ branching ratios are compared with those of Saito and Suzuki \cite{saito1992} in Table \ref{table_ions} for two energy regions.  The 660--670 eV region is below the 3d$_{5/2}$ - 6p Rydberg resonance at 674 eV, which is the lowest excitation energy of a 3d electron.  The ion branching ratios in that region are determined by photoionization of the 5s and 5p valence electrons and populate lower charge states compared with photon energies that excite 3d electrons.  Small variations of the branching ratios are observed in Figs. \ref{Xe_BR_1} and \ref{Xe_BR_2} from excitation of the 3d$_{5/2}$ - 6p Rydberg state at 674 eV, but the ion yields are clearly dominated by the 3d - $\epsilon$f shape resonances.

Table \ref{table_ions} also compares measurements made in the 695--730 eV range above the 3d$_{5/2}$ and  3d$_{3/2}$ binding energies.  Ionization of a 3d electron triggers a vacancy cascade, a series of Auger electron transitions through alternative pathways that lead to the observed range of charge states.\cite{jonauskas2003,partanen2005,viefhaus2005,suzuki2011}  The de-excitation pathways have been calculated using codes for the electronic structures of the hole states and Auger transition rates.\cite{jonauskas2003,partanen2005}  Corresponding calculations for core-hole decays in molecules are more challenging due to their non-spherical, multi-site structures.\cite{agren1992,tarantelli1995}  However, the atomic measurements and calculations provide guidance to understanding the first decay steps of Xe 3d holes in both Xe and XeF$_2$.  Jonauskas \textit{et al.}\cite{jonauskas2003} reported that 90$\%$ of 3d holes decay to double-hole states in the \textit{N} shell, i.e., 4d$^{-2}$, 4p$^{-1}$4d$^{-1}$, 4s$^{-1}$4d$^{-1}$, 4p$^{-2}$, and 4s$^{-1}$4p$^{-1}$.  Those holes are still localized.  Only 4$\%$ of 3d holes decay to states with 5s or 5p holes, i.e., 4d$^{-1}$5p$^{-1}$ and 4d$^{-1}$5s$^{-1}$.  It is not until the second Auger decay step that a significant fraction of 5p holes are produced.\cite{jonauskas2003}  In XeF$_2$, the Xe 5p electrons combine with F 2p electrons to form delocalized valence MOs.\cite{basch1971}  By comparison with the stepwise 3d hole decays in atomic Xe, we infer that in XeF$_2$ delocalized electrons participate in the second Auger decay steps of a Xe 3d hole and spread charge to the F sites.

\begin{widetext}
\begin{table}
\caption[]{\label{table_ions} Branching ratios for Xe ion charge states from Saito and Suzuki \cite{saito1992} at 670 eV and 700 eV compared with the present results for Xe averaged over 660--670 eV and 695--730 eV.  The last column shows the present results for Xe ion charge states from XeF$_2$ averaged over 695--730 eV.}
\label{Tab:charge-ratios}
\begin{ruledtabular}
\begin{tabular}{cccccc}\
ion & 670 eV$^a$ & 660--670 eV$^b$ & 700 eV$^a$ & 695--730 eV$^b$ & XeF$_2$: 695--730 eV$^b$\\ \hline
Xe$^+$ & 3.5 & 4.9 $\pm$ 0.2 & 0.7 & 0.7 $\pm$ 0.1 & 19.4 $\pm$ 0.1\\
Xe$^{2+}$ & 29 & 32.4 $\pm$ 0.3 & 4.0 & 5.5 $\pm$ 0.1 & 35.7 $\pm$ 0.1\\
Xe$^{3+}$ & 25 & 23.3 $\pm$ 0.2 & 7.5 & 7.8 $\pm$ 0.1 & 25.6 $\pm$ 0.1\\
Xe$^{4+}$ & 30 & 26.4 $\pm$ 0.3 & 41 & 39.8 $\pm$ 0.2 & 12.1 $\pm$ 0.1\\
Xe$^{5+}$ & 9.0 & 7.8 $\pm$ 0.1 & 29 & 27.9 $\pm$ 0.2 & 4.2 $\pm$ 0.1\\
Xe$^{6+}$ & 3.0 & 3.2 $\pm$ 0.1 & 15 & 15.0 $\pm$ 0.1 & 1.7 $\pm$ 0.1\\
Xe$^{7+}$ & 0.5 & 1.3 $\pm$ 0.1 & 2.5 & 2.8 $\pm$ 0.1 & 1.3 $\pm$ 0.1\\
Xe$^{8+}$ & 0.1 & 0.6 $\pm$ 0.1 & 0.3 & 0.5 $\pm$ 0.1 & --\\
Xe$^{9+}$ & -- & 0.1 $\pm$ 0.1 & -- & -- & --\\
\end{tabular}
\end{ruledtabular}
$^a$Ref. \citenum{saito1992}, $^b$This work.
\end{table}
\end{widetext}

\subsection{XeF$_2$ ion yields}
The Xe 3d$_{5/2}$ - $\epsilon$f and Xe 3d$_{3/2}$ - $\epsilon$f shape resonances dominate the total ion yield scans of Xe and XeF$_2$ plotted in Fig. \ref{Xe_XeF2_XS}.  Hartree-Fock-Slater calculations of the Xe 3d and F 1s photoionization cross sections averaged over 700--750 eV give a Xe 3d cross section 7$\times$ larger than twice the F 1s cross section.\cite{yeh1985}  The only clear signature of F 1s excitation is at the 683 eV LUMO resonance that also has a contribution from Xe 3d$_{3/2}$ excitation (see Tables \ref{Tab:Lan_Results1}--\ref{Tab:energetics}).  Away from the 683 eV resonance, Xe 3d core hole decay is the dominant contributor to the ion yields of XeF$_2$.  Previous work on atomic Xe suggests a model for XeF$_2$ in which a Xe 3d hole decays to two 4s, 4p, or 4d holes that are localized on Xe,\cite{jonauskas2003} followed by Auger transitions involving delocalized Xe 5p + F 2p MOs that spread charge to the F sites.  The shift to lower Xe$^{q+}$ charge states in going from the atom to the molecule shown in Fig. \ref{Xe_XeF2_iTOF} is quantified in Table \ref{Tab:charge-ratios} by comparing measurements over 695--730 eV.  With charge spread across the F and Xe sites, the system Coulomb explodes.  Energetic fragmentation is apparent in the ion spectra of F$^+$ and F$^{2+}$ plotted in Fig. \ref{F_F2_iTOF}.  We explained in Section \ref{methods} that our ion spectrometer has relatively low collection efficiencies for energetic F$^+$ and F$^{2+}$ ions but good sensitivity for Xe$^{q+}$ ions.  We can therefore compare Xe$^{q+}$ ion yields from the atom and molecule.  In addition, although the measured F$^+$ and F$^{2+}$ yields are small, their variations with photon energy are informative.

\begin{figure}
\begin{center}
\includegraphics [trim=5cm 7cm 7cm 3cm,clip,width=8cm]{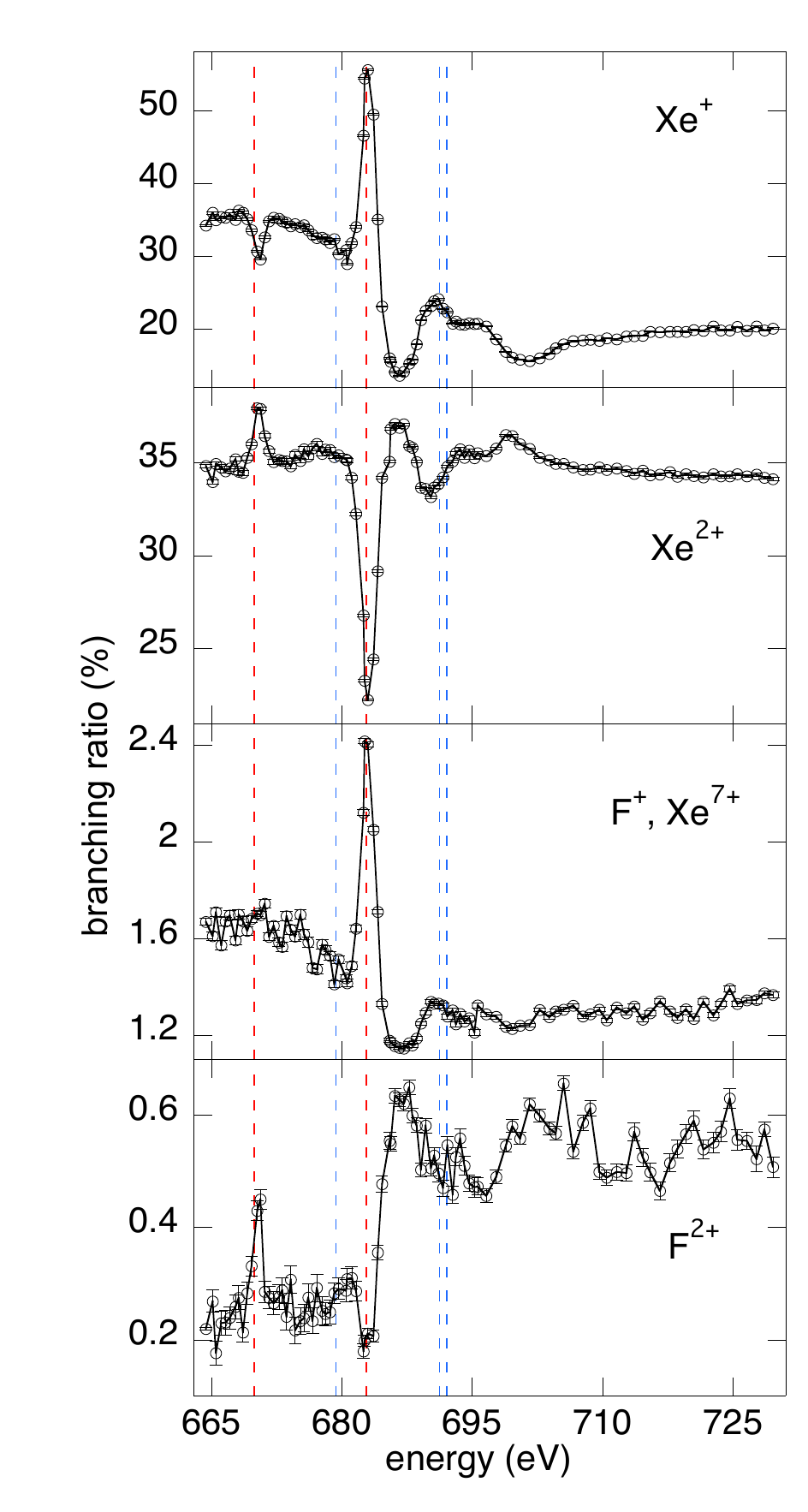}
\caption{\small Branching ratios for Xe$^+$, Xe$^{2+}$, F$^+$/Xe$^{7+}$, and F$^{2+}$ from photoionization of XeF$_2$.  The blue vertical lines mark the Xe 3d$_{5/2}$ (679.31 eV), Xe 3d$_{3/2}$ (692.09 eV), and F 1s (691.23 eV) binding energies.  The red vertical lines mark the Xe 3d$_{5/2}$ - LUMO (669.9 eV) and Xe 3d$_{3/2}$,F 1s - LUMO (682.8 eV) resonances.}
\label{XeF2_BR_1}
\end{center}
\end{figure}

\begin{figure}
\begin{center}
\includegraphics [trim=5cm 7cm 7cm 3cm,clip,width=8cm]{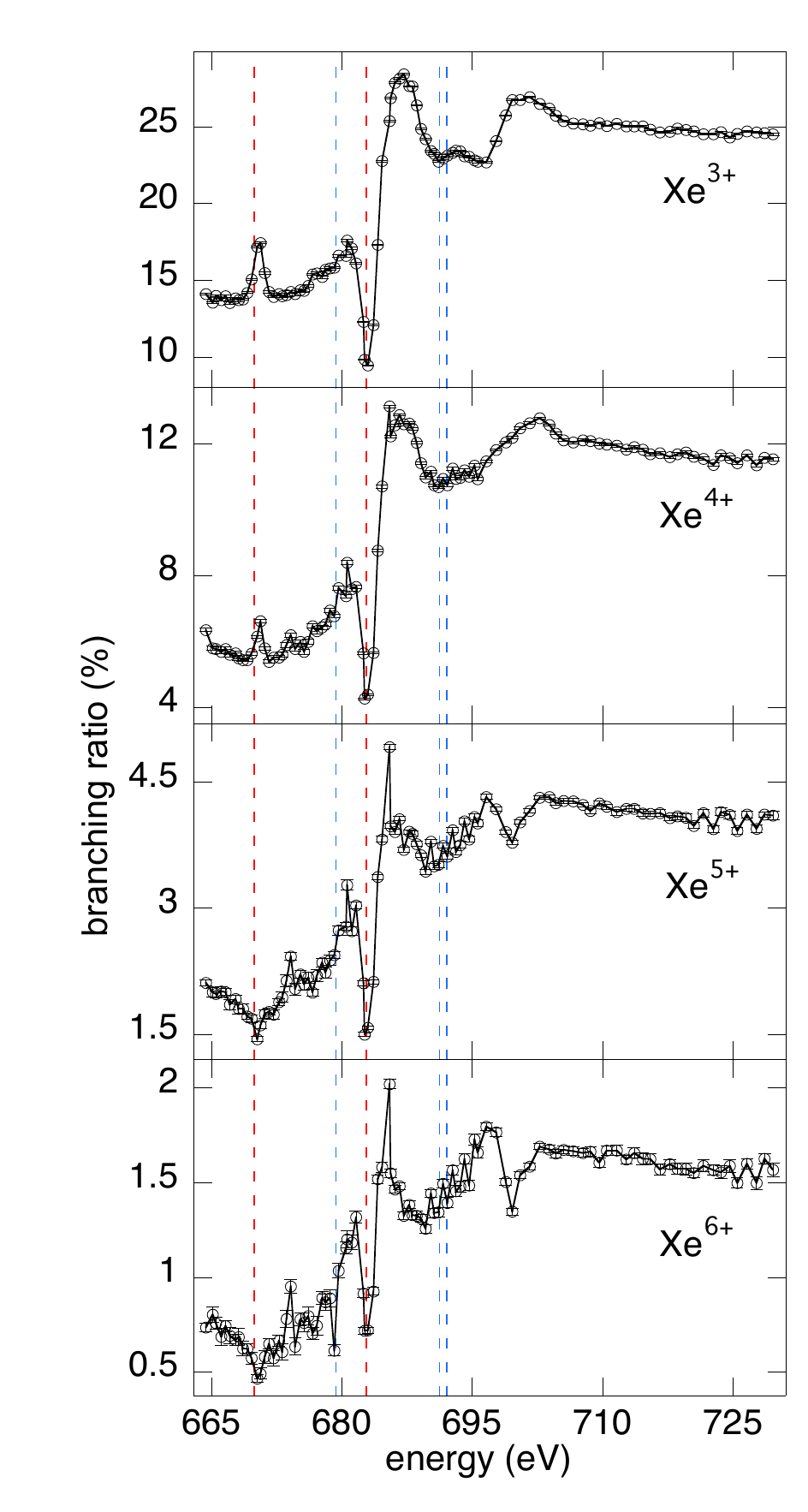}
\caption{\small Branching ratios for Xe$^{q+}$ (q = 3--6) from photoionization of XeF$_2$.  The blue vertical lines mark the Xe 3d$_{5/2}$ (679.31 eV), Xe 3d$_{3/2}$ (692.09 eV), and F 1s (691.23 eV) binding energies.  The red vertical lines mark the Xe 3d$_{5/2}$ - LUMO (669.9 eV) and Xe 3d$_{3/2}$,F 1s - LUMO (682.8 eV) resonances.}
\label{XeF2_BR_2}
\end{center}
\end{figure}

\begin{figure}
\begin{center}
\includegraphics [trim=5cm 7cm 7cm 3cm,clip,width=8cm]{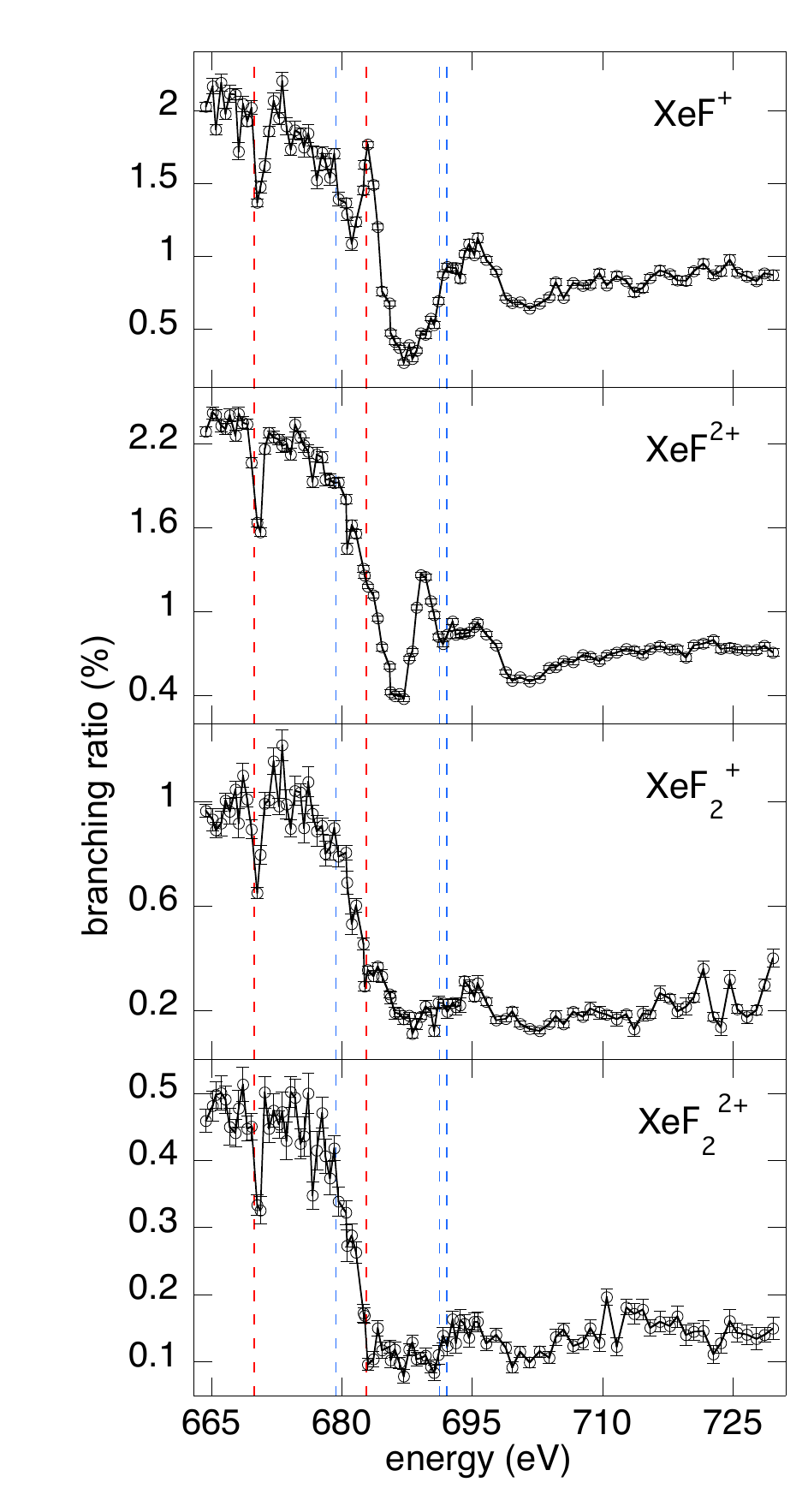}
\caption{\small Branching ratios for XeF$^+$, XeF$^{2+}$, XeF$_2^+$ and XeF$_2^{2+}$ from photoionization of XeF$_2$.  The blue vertical lines mark the Xe 3d$_{5/2}$ (679.31 eV), Xe 3d$_{3/2}$ (692.09 eV), and F 1s (691.23 eV) binding energies.  The red vertical lines mark the Xe 3d$_{5/2}$ - LUMO (669.9 eV) and Xe 3d$_{3/2}$,F 1s - LUMO (682.8 eV) resonances.}
\label{XeF2_BR_3}
\end{center}
\end{figure}

Ion spectra of XeF$_2$ were recorded over 665--730 eV and the counts under each peak summed and divided by the total counts to determine branching ratios.  No corrections were made for the low collection efficiencies of F$^+$ and F$^{2+}$.  The variations with energy are plotted in Figs. \ref{XeF2_BR_1}--\ref{XeF2_BR_3}.  Features are observed in response to the Xe 3d$_{5/2}$ - LUMO resonance near 670 eV, at the unresolved Xe 3d$_{3/2}$ - LUMO and F 1s - LUMO resonances near 683 eV, and from the two Xe 3d - $\epsilon$f shape resonances with maxima at 688 and 700 eV.

The first excitation of a Xe 3d$_{5/2}$ electron occurs at the 670 eV LUMO resonance that is located 9.4 eV below the 3d$_{5/2}$ ionization energy.  The calculations of Basch\cite{basch1971} characterize the LUMO as an anti-bonding MO with Mulliken populations of 68$\%$ Xe 5p, 31$\%$ F 2p, and 1$\%$ F 2s.  We expect the LUMO electron to act as a spectator in the first Auger decay step of the 3d$_{5/2}$ hole to two holes in the Xe \textit{N} shell but to participate in subsequent decay steps.  Our branching ratio measurements show decreases of Xe$^+$, Xe$^{5+}$, Xe$^{6+}$, XeF$_2^+$, XeF$_2^{2+}$, XeF$^+$, and XeF$^{2+}$.  The branching ratios for Xe$^{2+}$, Xe$^{3+}$, Xe$^{4+}$, and F$^{2+}$ increase, while the overlapped F$^+$ and Xe$^{7+}$ ions shows no response.

\begin{figure}
\begin{center}
\includegraphics [trim= 6cm 9cm 7cm 10cm,clip,width=8cm]{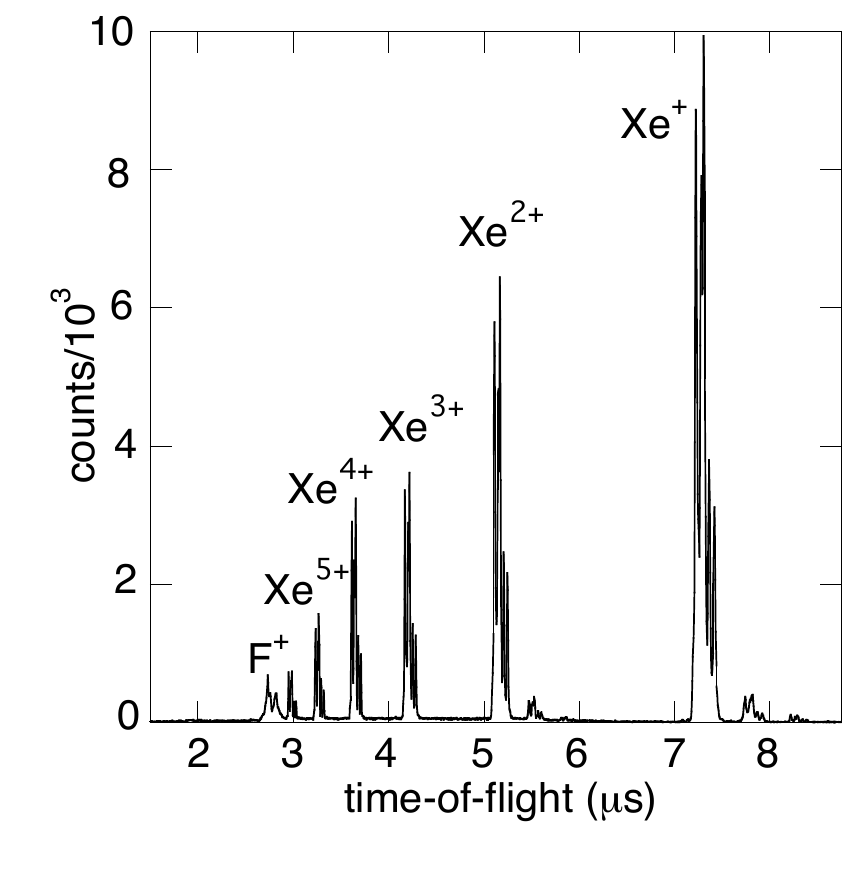}
\caption{\small Ion spectrum of XeF$_2$ recorded on the Xe 3d$_{3/2}$, F 1s - LUMO ($\sim$683 eV) resonance.  The relative yields of Xe$^+$ and F$^+$ fragment ions increase on resonance.  The splitting of the F$^+$ peak corresponds to $\sim$3 eV kinetic energy.}
\label{XeF2_iTOF_onres}
\end{center}
\end{figure}

The most dramatic variations of the branching ratios occur at the 683 eV resonance where the LUMO is accessed by both F 1s and Xe 3d$_{3/2}$ electrons.  The Xe$^+$, F$^+$, and XeF$^+$ ratios increase while the Xe$^{q+}$ (q = 2--6) and F$^{2+}$ ratios drop.  The ion spectrum recorded on resonance is plotted in Fig. \ref{XeF2_iTOF_onres} and shows enhanced production of Xe$^+$ and F$^+$ compared with Fig. \ref{Xe_XeF2_iTOF}.  In comparison with Fig. \ref{F_F2_iTOF}(a), the F$^+$ ions are much stronger on resonance than the Xe$^{7+}$ component.  This is consistent with the calculated oscillator strengths in Table \ref{Tab:Lan_Results2} that find the F 1s - LUMO resonance to be $\sim20\times$ stronger than the Xe 3d$_{3/2}$ - LUMO excitation.\cite{footnote1}  The F 1s - LUMO transition moment is parallel to the molecular axis, and the photon beam is highly linearly polarized (0.996 $\pm$ 0.003).\cite{whitfield2011}  Resonant excitation selectively excites molecules aligned with the polarization direction.  The ion spectrometer axis is aligned with the polarization, which contributes to the measured enhancement of the F$^+$ yield.  The splitting of the F$^+$ peak is smaller on resonance and corresponds to a fragmentation energy of only $\sim$3 eV.  These results distinguish F 1s excitation and decay from Xe 3d excitation and decay that dominates at other energies.  F 1s photoionization of several small molecules (SF$_6$, SiF$_4$, BF$_3$, CF$_4$) primarily produces singly- and doubly-charged fragment ions.\cite{piancastelli2005,piancastelli2008,guillemin2010jpb,guillemin2010pra} This is consistent with a simple picture of F 1s photoionization followed by a single Auger decay step that produces low total charge states.  With low total charge on the molecular ion, the kinetic energies of the fragment ions are smaller than for decays of Xe 3d holes.

The Xe 3d shape resonances dominate at photon energies above the 683 eV resonance.  The branching ratios of higher Xe$^{q+}$ charge states (q = 3--6) and F$^{2+}$ increase while the ratios of lower charge states (Xe$^+$, XeF$_2^+$, XeF$_2^{2+}$, XeF$^+$, and XeF$^{2+}$) decrease.  The F$^+$ yield in this energy range is unclear due to overlapping with Xe$^{7+}$.  However, enhancement of the F$^{2+}$ yield and the high kinetic energies of F$^+$ and F$^{2+}$ estimated from the ion spectra in Fig. \ref{F_F2_iTOF} are characteristic of Xe 3d hole decay in XeF$_2$.

\section{CONCLUSION}
\label{conclusion}

Inner-shell photoionization and core hole decay of Xe 3d$_{5/2}$, Xe 3d$_{3/2}$, and F 1s subshells of Xe and XeF$_2$ are compared over 660--740 eV to explore effects of the F ligands.  The total and partial ion yields vary with photon energy in response to Xe 3d$_{5/2}$ - $\epsilon$f and Xe 3d$_{3/2}$ - $\epsilon$f shape resonances in both the atom and molecule as well as excitation of the XeF$_2$ LUMO from Xe 3d and F 1s subshells. Coupled-cluster calculations accounting for relativistic effects give accurate values for ionization energies and resonance energies, and calculated oscillator strengths have helped to guide interpretations of measurements at the LUMO resonances.

We use the stepwise Auger decay model for 3d holes in atomic Xe\cite{jonauskas2003} to propose a corresponding model for XeF$_2$.  The first Auger decay step produces two holes in the Xe \textit{N} shell that are localized on Xe, but the following Auger transitions spread charge across the three atomic sites.  The system Coulomb explodes, producing Xe$^{q+}$ (q = 1--6) and energetic F$^+$ and F$^{2+}$ ions.  The molecule responds differently to excitation of the F 1s - LUMO resonance.  The relative yields of F$^+$ and Xe$^+$ increase, the F$^{2+}$ and Xe$^{q+}$ (q = 2--6) yields decrease, and the fragmentation energy of F$^+$ ions is relatively small. Therefore, XeF$_2$ molecule is very exceptional in the photon-energy range studied. First, the molecular core-hole decay can be compared with Xe atom, and second, either an electron from the F site or the Xe site can be excited/ionized giving rise to a totally different core-hole decay. 

Additional measurements are suggested by the results.  Momentum-resolved, multiple-ion-coincidence measurements with high-resolution synchrotron radiation can directly correlate the F and Xe fragment ions and their energies.\cite{pesic2008,laksman2013}  Core-hole decay and molecular fragmentation proceed on the femtosecond time scale, which is much faster than synchrotron x-ray pulses.  However, the intense femtosecond x-ray pulses produced at XFELs
can be used for time-resolved measurements.\cite{erk2014}  New capabilities are being developed at XFELs, such as using two femtosecond x-ray pulses with different colors and variable time delay.\cite{lutman2013}  The first pulse can trigger a core hole decay process in a molecule while the second pulse probes the dynamics.\cite{picon2014}

Core hole decay dynamics in molecules is a challenging problem for theory and computational methods. Treatments require a proper description of orbital relaxation and electron correlation. Moreover, inner-shell excitations induce multiply excited configurations due to ``shake-up'' processes making the convergence of excited state calculations very difficult. The lineshape of inner-shell excitations can be elucidated by calculating the Auger decay of the inner-excited states. Calculations of Auger decays need an accurate description of the continuum for the emitted Auger electron. This is quite challenging to implement within commonly used bases, such as Gaussian bases, and some methods have been implemented to address this problem \cite{Carravetta1987,Demekhin2011}. In our particular case, the decay from Xe 3d holes in XeF$_2$, two or more Auger transitions need to be treated while the molecular geometry expands and fragments, involving also an important motion of the nuclei. 

\begin{acknowledgments}
We thank the staff of the Synchrotron Radiation Center for technical support, and we thank Dr. Jeff Hammond for discussions of calculational methods.  The SRC was previously supported by the NSF and the University of Wisconsin--Madison.  SHS, AP, and CSL were supported by the U.S. Department of Energy, Office of Science, Chemical Sciences, Geosciences, and Biosciences Division under contract no. DE-AC02-06CH11357. LC and JFS are grateful to the US National Science Foundation (Grant CHE-1361031) and the Robert A. Welch Foundation of Houston, TX (Grant F-1283).

\end{acknowledgments}


\begin{thebibliography}{99}
\bibitem{arnoldi}
	We note that the unsymmetric Lanczos algorithm is computationally advantageous, as no storation of the trial functions
	is needed for energy calculations. In the present work, we have adopted the Arnoldi algorithm to expedite the implementation.
\bibitem{footnote2}
	We have performed a set of calculations with the cutoff threshold for the virtual orbitals being 20 a.u., 50 a.u., and 100 a.u. to study the convergence of the results. It is shown that the errors due to a virtual cutoff at 100 a.u. is less than 0.1 eV.
\bibitem{footnote1}
	The values of the oscillator strength for Xe 3d$_{3/2}$ to LUMO and 3d$_{5/2}$ to LUMO transitions are obtained to be 0.0029 and 0.0044, respectively, by transforming the spin-free values in Table \ref{Tab:Lan_Results2}.
\end{thebibliography}

\end{document}